\documentstyle [aps,preprint,prl] {revtex}

\begin{document}
\draft
\preprint{nd95-02}

\title{Boundary-induced wavelength selection in a one-dimensional
pattern-forming system}

\author{Samuel S. Mao and John R. de Bruyn}
\address{Department of Physics\\
Memorial University of Newfoundland\\
St. John's, Newfoundland, Canada A1B 3X7}

\author{Stephen W. Morris}
\address{Department of Physics and Erindale College\\
University of Toronto\\
60 St. George St.,\\
Toronto, Ontario, Canada M5S 1A7}

\date{\today}
\maketitle

\begin{abstract}

We have measured the stability boundary for steady electrically-driven
convective flow in thin, freely suspended films of Smectic-A liquid
crystal. The thinness and layered structure of the films supress two-
and three-dimensional instabilities of the convection pattern. As the
voltage applied across the film, or the length of the film, is varied,
convective vortices are created or destroyed to keep the wave number
of the pattern within a stable range. The range of stable wave numbers
increases linearly with the dimensionless control parameter
$\epsilon$, for small $\epsilon$, and the vortices always appear and
disappear at the ends of the film. These results are consistent with a
mechanism for boundary-induced wavelength selection proposed by Cross
{\it et al.} [Phys.  Rev. Lett. {\bf 45}, 898 (1980)].

\end{abstract}


The formation and dynamics of patterns in nonequilibrium nonlinear
systems has been the subject of much recent study \cite{ch93}.  The
general behavior of fully three-dimensional systems can be quite
complicated, so there is substantial interest in systems which develop
simpler, one-dimensional patterns which can be subjected to detailed
experimental and theoretical study. Many such systems have been
studied, of which Rayleigh-B\'enard convection (RBC), in which a
pattern of straight convective rolls develops in a thin layer of fluid
heated from below, is perhaps the best known \cite{ch93,a89}. Although
the fluid layer is three-dimensional, the flow {\it pattern} just
above onset is characterized by a one-dimensional wave vector $k$ and
one may speak of a one-dimensional pattern.  An important issue in the
study of patterns is that of wave number selection \cite{ch93}.
Typically, beyond the transition at which a pattern first appears, the
base state will be unstable to perturbations within a band of wave
numbers.  The range of wave numbers over which the pattern is stable
is limited to a narrower band of selected wave numbers by various
instabilities. In the case of RBC, many of these instabilities involve
perturbations in directions other than that of the original wave
vector and lead to two- or three-dimensional flow patterns \cite{b84}.
These can be partly supressed by confining the fluid layer to a
container which is small in the direction parallel to the roll axis,
although the flow velocity field will always be three-dimensional.
Some wave number selection mechanisms are purely one dimensional: The
Eckhaus instability \cite{e65,m90} is a long-wavelength phase
instability which results in the creation or loss of pattern units
({\it i.e.,} of convective roll pairs in the case of RBC) in the interior of
the experimental cell. Slow spatial variation of the control parameter
through the value at which the pattern appears will select a unique
wave number \cite{kbbc82,cda83}. Another one-dimensional wave number
selection mechanism results from the finite size of the experimental
system \cite{cdhs}.  For the case of RBC, the boundary conditions on
the flow at end walls perpendicular to $k$ are predicted to allow the
creation or destruction of rolls at the end walls, resulting in a
change of wave number. Numerical simulations of RBC have shown roll
creation and loss at end walls\cite{mhl86,abn87}, but this mechanism
has not been unambiguously observed in RBC experiments\cite{mhlpp84}
due to the three-dimensional nature of the velocity field \cite{ch93}.

In this Letter we report measurements of the stability boundary of
electrically-driven convective flow in very thin, freely suspended
films of smectic-A liquid crystal \cite{mdm90,mdm91b,m91}.
Smectics-A have a layered structure \cite{d79} which makes it possible
to make centimeter-size films, fractions of a micron thick, which are
completely uniform in thickness. Flow between the layers is very
difficult, and a film behaves as a two-dimensional isotropic fluid
\cite{mdm91b}. When a voltage $V$ is applied across the film, a convection
pattern in the form of a one-dimensional array of counter-rotating
vortices appears at a well-defined critical voltage $V_c$
\cite{mdm90}. The dimension of the smectic films in the direction of
the axis of the convective vortices is extremely small. As a result,
instabilities involving a bending or reorientation of the
vortex axis are impossible, and we expect only one-dimensional processes
to be involved in wave number selection. Indeed, we find that
wave number adjustments occur via the creation or loss of vortices at
the ends of the film. The shape of the stability boundary we measure
is consistent with that expected from the boundary-induced wave number
selection mechanism of Cross {\it et al.}
\cite{cdhs}

Our experimental apparatus is similar to that used previously
\cite{mdm90,mdm91b,m91}. The long sides of
the smectic-A film were supported by two 23 $\mu$m-diameter tungsten
wire electrodes.  The ends of the film were supported by thin plastic
wipers which rested on the wires.  The separation of the electrodes
$d$ was adjustable; for the work reported here $0.66$ mm $<d< 2$ mm.
One of the end wipers could be driven with a motorized micrometer,
allowing variation of the film length $l$ in the range $ 0 < l < 30 $
mm. Films were made by bringing the two wipers together, placing a
small amount of liquid crystal \cite{8cb} on the place where they
joined, and then slowly drawing them apart.  Films consist of an
integer number of smectic layers (one layer $=$ 3.16 nm) and can be
made uniformly thick.  They can maintain their uniform thickness even
in the presence of strong convection, and even when the length of the
film is changed, as it is easier for the film to exchange material
with the electrodes or the wipers than to form a partial smectic
layer.  A voltage is applied between the two electrodes, and steady
convection starts at $V = V_c$ and persists up to a certain voltage,
beyond which the flow becomes unsteady. The film holder was
temperature controlled to $\pm 0.1$ $^\circ$C over a given run.  All
runs were performed at temperatures in the range $25
\pm 1$ $^\circ$C, well below the smectic-A--nematic transition at 33
$^\circ$C.

The smectic films were viewed through a microscope with a color ccd
video camera. A small amount of incense smoke was admitted into the
experimental housing and some smoke particles settled on the film.
These particles were advected by flow in the film. Their motion was
followed by shining a collimated beam of white light onto the film
from below, and observing the scattered light.

The drawing procedure produces films of various uniform thicknesses $s$.
Since $V_c$ and thus the dimensionless control parameter \cite{mdm91b}
$\epsilon = (V/V_c)^2 -1 $ depend on $s$, it must be accurately
measured. We determined the film thickness in two ways. Since $s$ is
on the order of a wavelength of visible light, the films show bright
interference colors when viewed in reflected white light. The color of
the film can be calculated in terms of the CIE chromaticity
diagram\cite{spsc87}, and with practice the thickness of films up to
about 100 layers thick can be determined to an accuracy of $\pm 2$
layers from their color.  We also determined the thickness with an
accuracy of $\pm 1$ layer from reflectivity measurements \cite{ra80}.
These two techniques together permitted unambiguous determination of
the film thickness to an accuracy of $\pm 1$ layer.

Our measurements of the stability boundary of the steady convective
state were made using two techniques. In the first, a stable pattern
was prepared by increasing the applied voltage to a chosen value above
the onset of convection. Patterns with wave numbers $k$ close to $k_c
$, the wave number at onset \cite{mdm90}, were easily obtained by
increasing $V$ slowly through the pattern onset, while a sudden jump
from below to substantially above $V_c$ resulted in a pattern with $k$
different from $k_c$. Roughly speaking, jumping to values of
$\epsilon$ in the range $1 \lesssim \epsilon \lesssim 4$ gave $k<
k_c$, while jumping to $\epsilon \gtrsim 4$ resulted in a pattern with
$k>k_c$.  The voltage was then increased or decreased in small steps
and the flow pattern monitored. The pattern required about 50 vortex
turnover times, corresponding to about 1 minute, to equilibrate after
a change in $V$.  Typically, varying the voltage eventually led to a
wave number adjustment through either the creation or the loss of a
vortex, which always occurred at the ends of the film, not in its
interior.

In the second type of measurement, a stable pattern was prepared as
above.  The film length was then changed at constant thickness by
slowly ($\approx 20$ $\mu$m/s) moving the motorized wiper. This
resulted in a stretching (for increasing $l$) or a compression (for
decreasing $l$) of the vortex pattern, and periodically led to the
creation or loss, respectively, of one or more vortices when $k$
reached the stability boundary.  As above, the
creation and destruction of vortices occured at the ends of the film
\cite{middle,tvf}.

Figure \ref{wave-v} shows the stability range of the steady vortex
pattern measured by varying the voltage across the film at fixed film
length. Data from thirty-six runs, using films with thickness between
2 and 45 layers, and with aspect ratios in the range $3<l/d < 15$,
were combined to produce the stability boundary in Fig. \ref{wave-v}.
The data shown represent the maximum and minimum values of $\epsilon $
at which a given wave number state was observed. For some films, wave
number changes occurred inside the plotted boundary. This may be due
to variations in the end conditions in different runs, as discussed
below. No systematic variation in the position of the boundary with
either film thickness or aspect ratio was detected.

Figure \ref{wave-l} shows the wavelength $\lambda$ of the pattern in
two experiments in which the film length was changed by moving the end
wiper. As $l$ is increased (solid circles), $\lambda$ increases, to
accommodate the change in length at fixed number of vortices.
Eventually $\lambda$ reaches a value above which the pattern is
unstable. At this point one or more new vortices form at the
end of the film, and the mean wavelength of the pattern
decreases back to a stable value.  When the film length is decreased
(open circles) the opposite process occurs, with a vortex disappearing
at the end of the film when $\lambda$ decreases below a
stability boundary. As shown in Fig. \ref{wave-l}, the appearance and
disappearance of vortices is always hysteretic.

The stability boundary determined from data such as that in Fig.
\ref{wave-l} is plotted in Fig. \ref{bdy-l}, which shows the range of
wave numbers observed at different values of $\epsilon$. Data from 20
runs with films having $s$ between 3 and 65 layers and $2<l/d<15 $ are
shown. Again there were no systematic variations in the boundary with
either film thickness or aspect ratio.

The stability boundaries for $\epsilon < 2.5$, measured with these two
techniques, are shown superimposed in Fig. \ref{both_bdys}. They are
consistent in their general shape over the whole range of
measurements, and for $k-k_c < 0$ and at low $\epsilon$, the two
boundaries agree very well.  The quantitative differences in other
parts of the diagram may be due to variations in the
conditions at the ends of the films, as the small amount of liquid crystal
remaining on the wipers after the film has been drawn is different in
every run.

Cross {\it et al.} have studied the effects of endwalls on wave number
selection in RBC \cite{cdhs}. The convective flow velocity must go to
zero near an end wall, and as a result the amplitude of the convection
pattern is weaker near the wall than in the bulk. In the theory of
Cross {\it et al.}, this implies that it is easier to create or
destroy convection rolls at the ends of the experimental cell than in
the bulk. They calculated the range of stable wave numbers for this
situation and found it to be linear, {\it i.e.,} $\epsilon_B \propto
|(k-k_c)/k_c| $, in contrast to the quadratic boundary found for the
Eckhaus instability\cite{ch93,e65,m90}.  Thus at small enough
$\epsilon$, this boundary-induced wave number selection mechanism will
result in a narrower stable band than would the Eckhaus instability.
The slope of the lower-$k$ boundary of the stable region must be
negative, but that of the higher-$k$ boundary can have either sign,
depending, for RBC, on the fluid properties and on the exact nature of
the boundary conditions at the end walls \cite{cdhs}.

For small $\epsilon$, the  measured stability boundary is linear with
negative slope for $k<k_c$.  The dotted curve plotted in Fig.
\ref{both_bdys} is a fit to both sets of data for $\epsilon<1$,
$k<k_c$ of the function $\epsilon = a(k-k_c)/k_c + b\left(
(k-k_c)/k_c\right )^2$ with $a$ and $b$ parameters; it describes the
data in this range well.  Furthermore, in our experiments vortices
form or vanish at the ends of the system.  These results are
consistent with what would be expected from the boundary-induced wave
number selection mechanism \cite{cdhs}, and
inconsistent with what would be expected from the Eckhaus instability
\cite{ch93,e65,m90,cda83}.  For $k>k_c$ the measured stability
boundary is less well-determined, but within our uncertainties it appears
more linear than parabolic, and so is also consistent with the
mechanism of Cross {\it et al.}\cite{cdhs}

Measurements of the velocity field in a convecting film indicate that
the amplitude of the convection decreases near the ends, as expected
from a no-slip boundary condition on the flow at the rigid end
walls\cite{m95}. However, the actual end conditions in our experiments
involve more than just the flow condition, as the electric field which
drives the flow will be modified by the presence of the ends of the
film, and possibly also by the conductivity of the excess liquid
crystal which remains on the wipers.  In the context of the theory of
Ref.
\cite{cdhs}, such details might be expected to affect the slopes of
the branches of the stability boundary, but not their linear form.

In summary, we have observed wave number selection due to end
conditions in electrically-driven convection in freely suspended
smectic films.  Because of the extreme two-dimensionality of the
films, effects due to three-dimensional flow, or to bending of the
convection roll axis, do not exist in this system. Wave number
adjustments occur via the creation or loss of vortices at the ends of
the film, and the band of stable wave numbers broadens linearly for
small $\epsilon$. These results are consistent with those expected
from the theory of Cross {\it et al.} \cite{cdhs}, and not with the
bulk Eckhaus mechanism.

J. de B. is grateful to M. Cross for a helpful discussion. This
research was supported by the Natural Sciences and Engineering
Research Council of Canada.

\begin{figure}

\caption{The stable wave number range for steady electically-driven
convection, measured by increasing (circles) or decreasing (triangles)
the applied voltage. }

\label{wave-v}
\end{figure}

\begin{figure}

\caption{The pattern wavelength as function of length  when the
film length was varied. The wavelength plotted is the mean over the
pattern, excluding the vortices at the ends of the film.  The solid and
open circles were obtained by increasing and decreasing $l$,
respectively. Arrows indicate wavelength changes caused by the
creation or loss of vortices. a) $s=50$ layers, $\epsilon = 1.0$;
Single vortices are gained or lost at the arrows.  b) $s = 25 $
layers, $\epsilon = 3.0$; two vortices are gained
simultaneously at the downward arrows; they are lost one-at-a-time at
the upward arrows. }

\label{wave-l}
\end{figure}

\begin{figure}

\caption{The stability boundary determined by varying the
film length at constant $\epsilon$, as in Fig.
\protect\ref{wave-l}. }

\label{bdy-l}
\end{figure}

\begin{figure}

\caption{The stability boundaries from Figs. \protect\ref{wave-v} and
\protect\ref{bdy-l} plotted together for low $\epsilon$.
The symbols correspond to those in Figs. \protect\ref{wave-v} and
\protect\ref{bdy-l}. The dotted line is a fit to the data for $\epsilon<1$,
$k<k_c$ to the  form $\epsilon = a(k-k_c)/k_c + b\left(
(k-k_c)/k_c\right )^2$ which gives $a= -2.2 \pm 0.5$ and $b = 3.2 \pm 2.3$.}

\label{both_bdys}
\end{figure}

\vfill\eject


\begin{references}
\bibitem{8cb} The material used was
4,4$'$-n-octylcyanobiphenyl (8CB) doped with $7.5 \pm 0.2$ mM/l
tetracyanoquinodimethan (TCNQ) to control the nature of the ionic
species in the liquid crystal.
\bibitem{middle} At higher voltages, close to the onset of the unsteady state,
we occasionally observed vortices created or destroyed
in the interior of the film. Away from that transition, the wavelength
adjustments always took place at the ends.
\end{references}
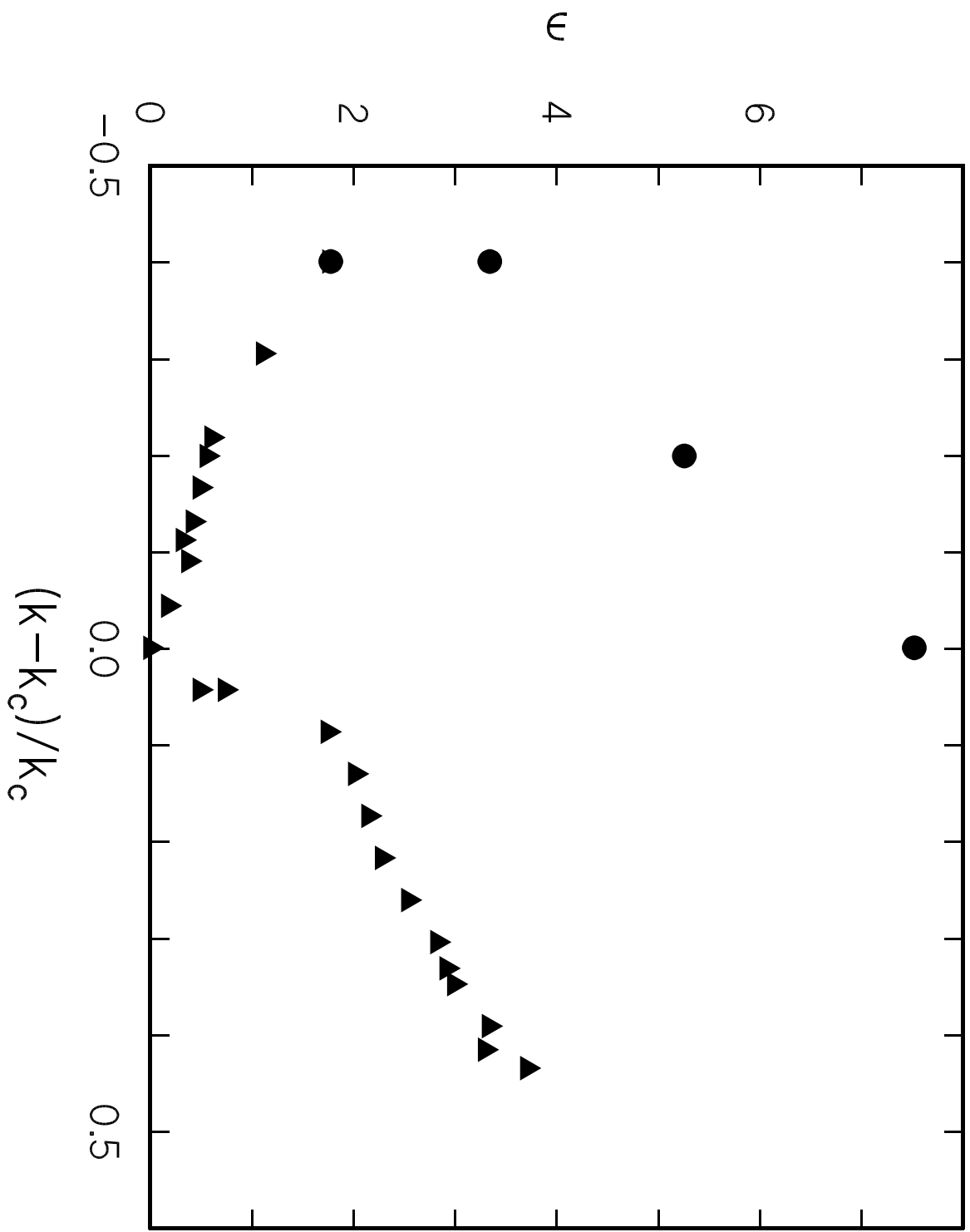
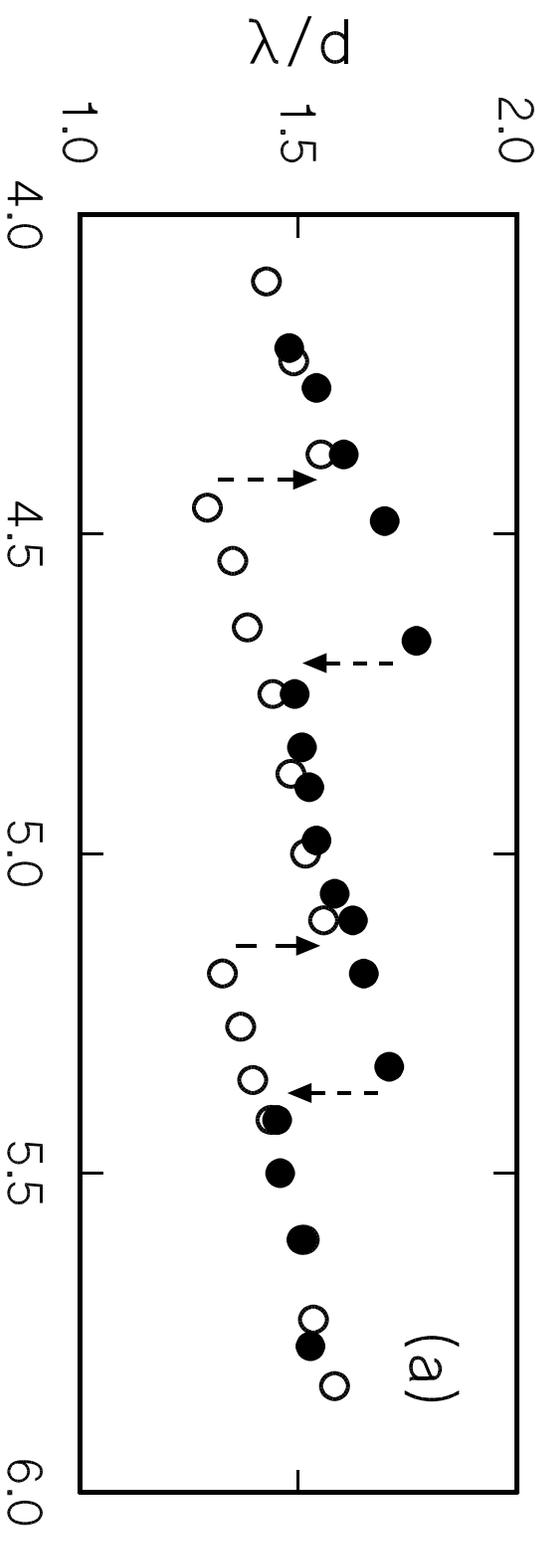
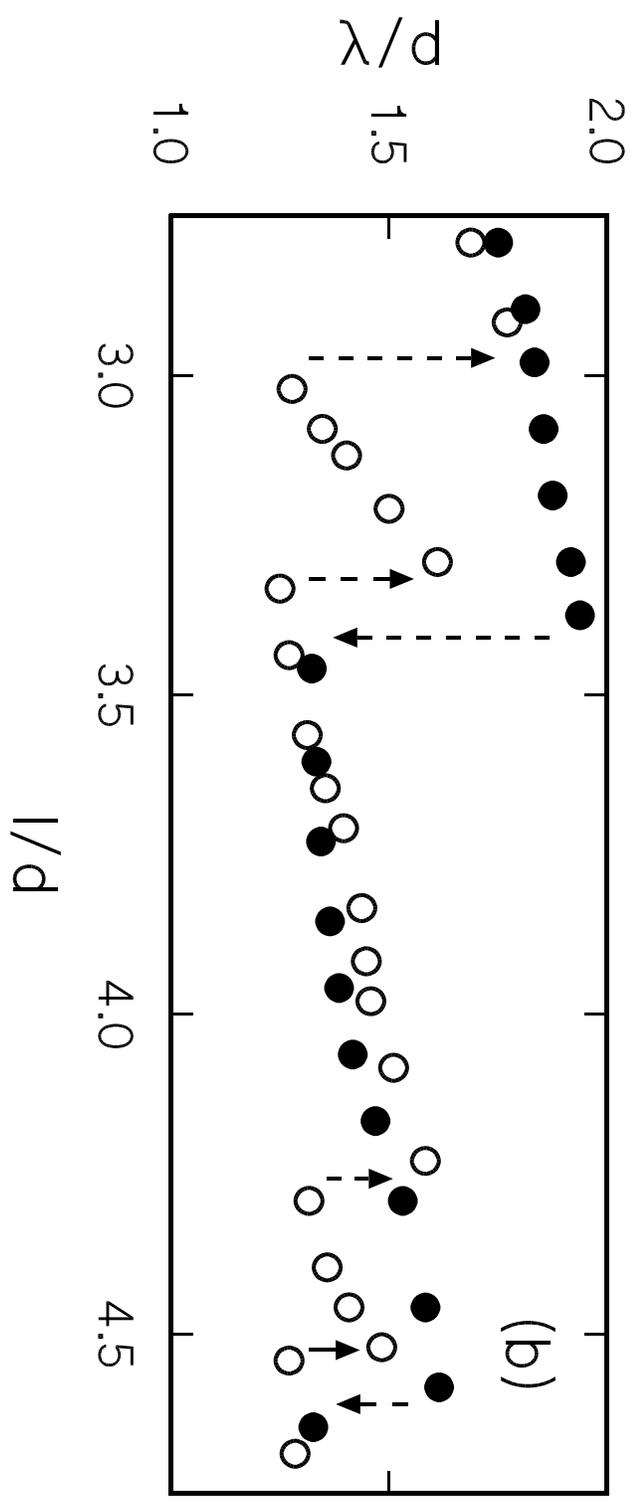
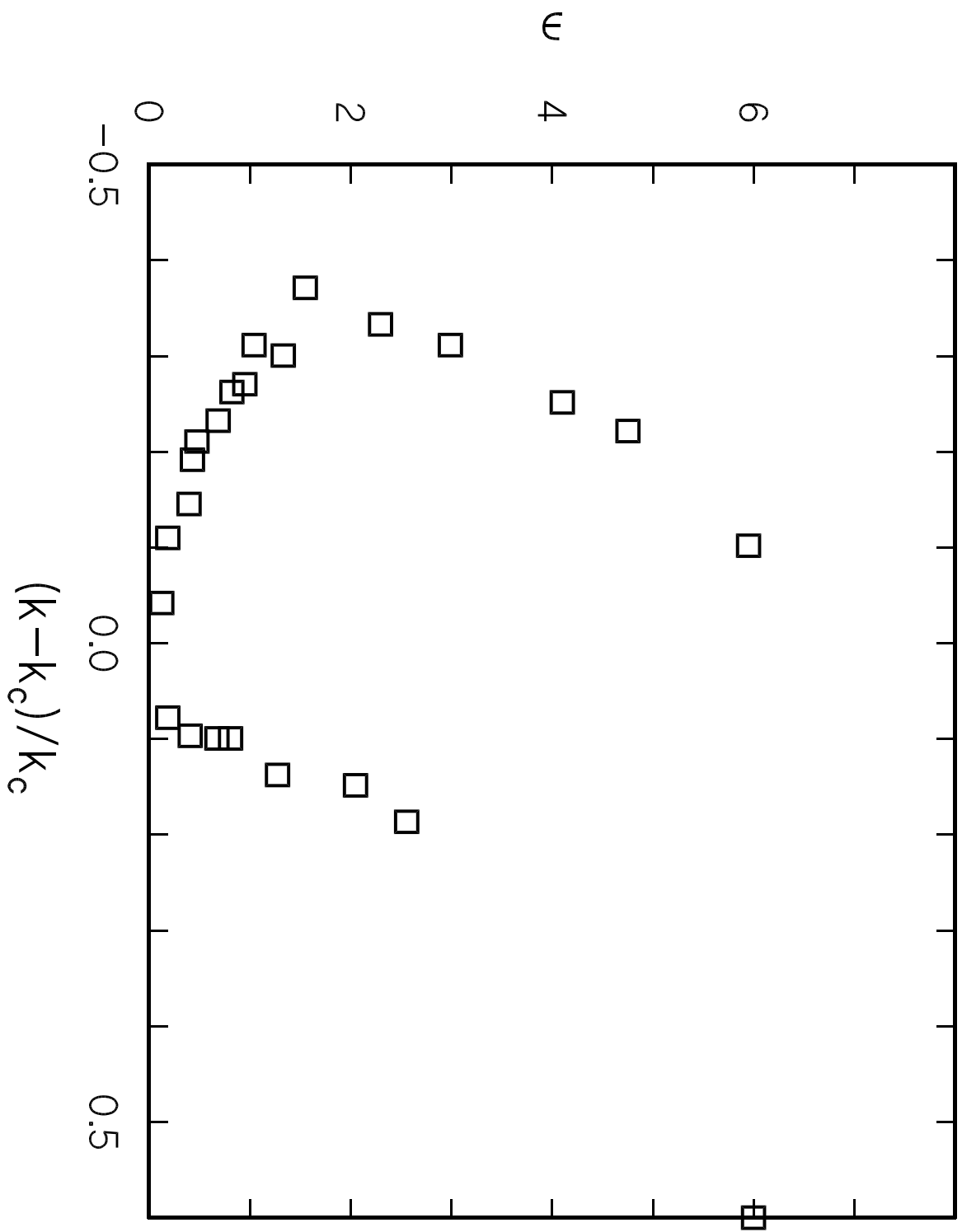
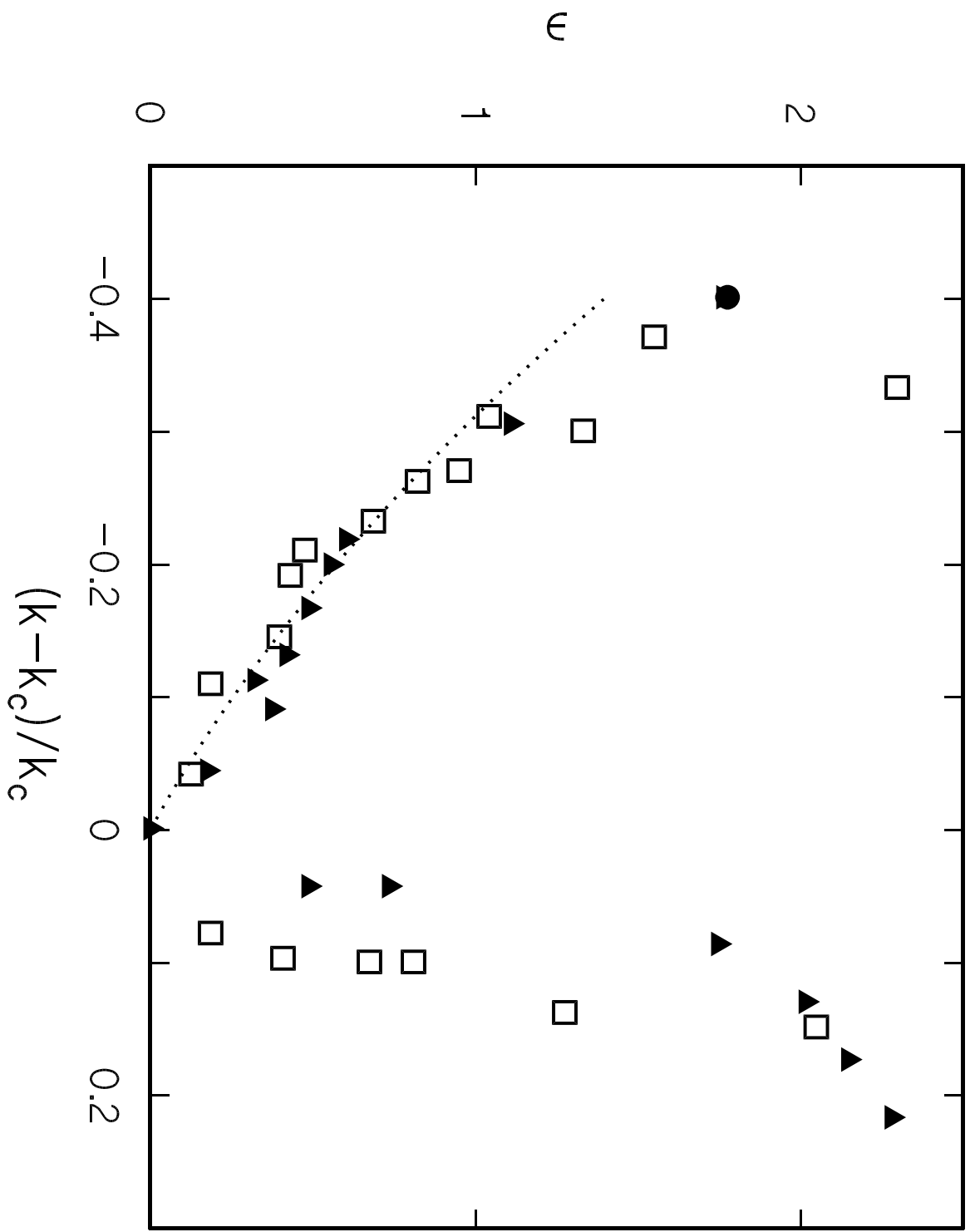
\end{document}